\documentclass[pra,twocolumn,showpacs]{revtex4}
\usepackage{epsfig}
\usepackage{graphicx}
\usepackage{amsmath}
\usepackage{natbib}
\newcommand{\bn}{\begin{eqnarray}}
\newcommand{\en}{\end{eqnarray}}
\newcommand{\eml}{\end{multline}}
\newcommand{\bml}{\begin{multline}}

\begin{document}

\title {Spin squeezing and Schr\"{o}dinger cat generation in atomic samples with Rydberg blockade}
 \author{Tom\'{a}$\check{{\rm s}}$ Opatrn\'{y}$^1$ and Klaus M{\o}lmer$^2$ }
 \affiliation{$^1$Optics Department, Faculty of Science, Palack\'{y} University, 17. Listopadu 12,
 77146 Olomouc, Czech Republic\\
 $^2$Lundbeck Foundation Theoretical Center for Quantum System Research, Department of Physics and Astronomy, University of Aarhus, DK-8000 Aarhus C, Denmark}

\date{\today }
\begin{abstract}
A scheme is proposed to prepare squeezed states and Schr\"{o}dinger cat-like states of the collective spin degrees of freedom associated with a pair of ground states in an atomic ensemble. The scheme uses an effective Jaynes-Cummings interaction which can be provided by excitation of the atoms to Rydberg states and an effective $J_x$ interaction implemented by a resonant Raman coupling between the atomic ground states. Both dynamical evolution with a constant Hamiltonian and with adiabatic variation of the two interaction terms  are studied. We show that by the application of further resonant laser fields, we can suppress non-adiabatic transfer under the time varying Hamiltonian and significantly speed up the evolution towards a maximally squeezed, $J_z=0$, collective spin state.
\end{abstract}
%

\maketitle

\section{Introduction}
Atomic ensembles hold great potential for precision metrology \cite{Wineland1994,Lloyd2006} as well as for quantum information processing \cite{Hammerer2010,Kozhekin2000,Julsgaard2004}, where one can take advantage of the discrete character of  spin states of an individual atom as well as of the quasi-continuous state structure of large atomic samples. In metrology, the precession of the atomic spin is used for time measurement in atomic frequency standards and  for high precision probing of, e.g., magnetic and electric fields. Engineering collective spin squeezed and entangled quantum states makes it possible to suppress noise and increase the measurement precision and sensitivity
\cite{Wineland1994,Kitagawa,Lloyd2006,Madsen2004,Pezze2009,Wasilewski2010}. The ladder of quantum states of the collective spin is formally compatible with the structure of quantized radiation modes, and the conventional atom-light interaction thus enables the construction of matter-light interfaces for quantum communication, where one may also benefit from ensembles being initially prepared in squeezed and entangled states \cite{Kuzmich2000}.

Atomic ensembles are readily manipulated with laser fields that drive transitions between internal atomic states. To produce states with entanglement between different atoms, however, more sophisticated interaction schemes are necessary. To prepare such states, one either needs to bring the ``nonclassicality'' from outside, e.g., by absorption of nonclassical states of light \cite{Kuzmich1997} or by the quantum back-action of an optical measurement of a collective atomic observable \cite{Julsgaard2004,Madsen2004,Wasilewski2010}, or one may induce suitable interatomic interactions within the ensemble. The former approaches have their limits as not all quantum states of the light field are easily available and the measurement sensitivity has to beat the usual shot noise limit; and to have sufficient atom-light coupling strength, they typically work only for very large samples.  The latter approach is not easy because atomic ensembles are typically dilute and the atoms move around randomly: the  interactions among the atoms, needed for quantum state preparation, are thus also a source of decoherence.

In this paper we propose to use the interaction between Rydberg excited states \cite{Gallagher} to correlate the atoms within an ensemble in a controllable way. There have been various proposals to use the Rydberg interaction for quantum state manipulations and for quantum information processing \cite{Jaksch2000,Lukin2001,Bouchoule2002,Fleischhauer2002,Saffman2005,Moller08,Zoller2009,Saffman2009,Saffman2010}. As in a number of previous proposals, we make use of the so-called blockade effect, where the excitation of one single atom shifts the energy and thus prevents the resonant excitation of other nearby atoms. This effect on a single pair of atoms has been observed and used to create entangled states and perform quantum gates \cite{Urban09,Gaetan09} and in large atomic samples it leads to a significant suppression of the excitation number and number fluctuations under resonant irradiation \cite{Low2009,Liebisch2005}. We assume an ensemble of atoms confined to a spatial volume so that all atoms are within the interaction range of the others. The quantized occupation of two stable atomic ground states is then represented by harmonic oscillator degrees of freedom while the restriction of the number of Rydberg excited atoms to zero and unity permits a mapping of this degree of freedom on an effective two-level system. Moreover, the coherent coupling of an atomic Rydberg and ground state has the form of the Jaynes Cummings (JC) model \cite{JaynesCummings1963} of the interaction between a two-level atom and the quantized radiation field in quantum optics.

We will investigate to what extent the well known theoretical potential to generate squeezed states \cite{Meystre1982,Banacloche93} and Schr\"{o}dinger cat states \cite{Buzek1992} of light applies also to the case of collective atomic spins. Apart from the formal analogy, which allows us to directly apply the results of the JC coupling, our scheme takes advantage of some special properties of the atomic oscillator system as compared to the quantized field: {\em (i)} the atoms populate two ground  states and they can be prepared in the spin-coherent state at the equator of the Bloch sphere where the binomial number statistics has smaller fluctuation than the corresponding coherent state of the harmonic oscillator with the same mean occupation, {\em (ii)} the JC interaction can  be switched on and off and both the intensity and phase of the JC Hamiltonian can be relatively easily varied in time, {\em (iii)} one can couple each of the low lying atomic levels to separate Rydberg state and thus eliminate some unwanted asymmetries generated in the standard JC scenario, and {\em (iv)} one can simultaneously act by resonant Raman fields to adjust the atomic ground state. There are of course numerous further practical differences between squeezing of light and atoms, both concerning the lifetime of the states produced and their possible applications.

The paper is organized as follows. In Sec.~\ref{sec-syst}, we introduce the physical system and our  notation. In Sec.~\ref{sec-dyn}, we present numerical and analytical results for dynamical squeezing with the effective JC interaction. In Sec.~\ref{sec-adiab}, we determine the eigenstates of a Hamiltonian with both Raman and JC coupling terms, we show that it adiabatically connects a spin coherent state with a maximally spin-squeezed state, and we show that application of judiciously chosen additional couplings suppresses non-adiabatic processes and permits rapid evolution along the desired eigenstates. In Sec.~\ref{sec-schroed} we analyze the production of Schr\"{o}dinger cat states by the JC interaction. In Sec.~\ref{sec-concl}, we summarize our results.


\section{System description}
\label{sec-syst}

\begin{figure}
\centerline{\epsfig{file=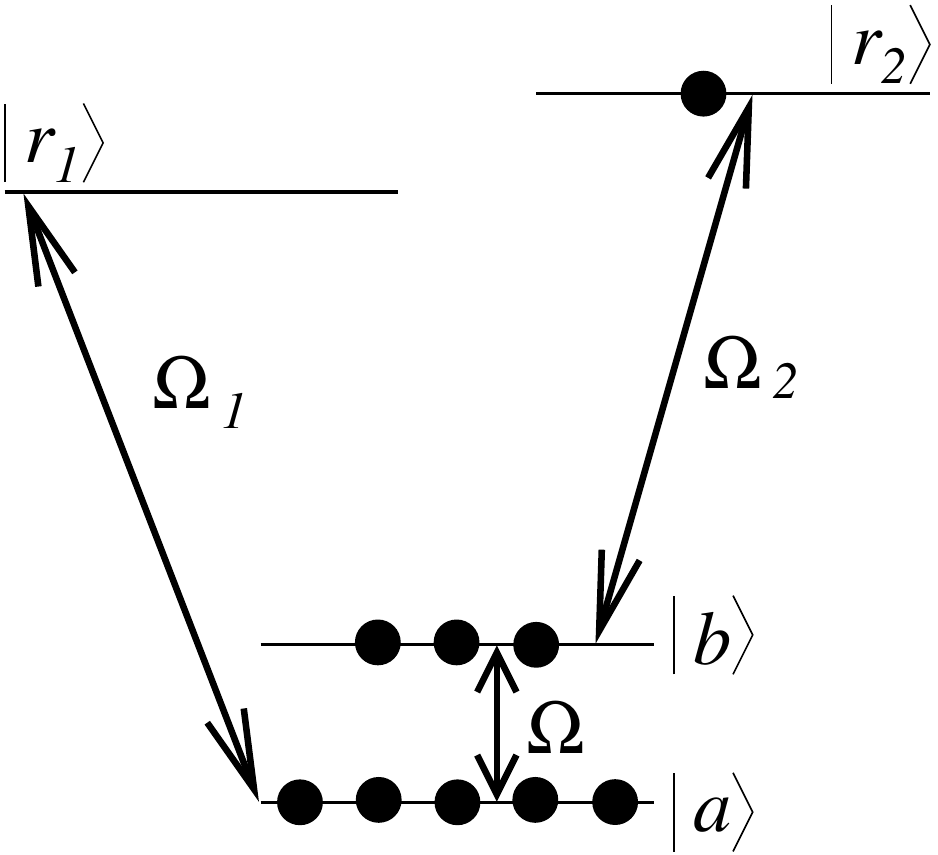,scale=0.4}}
\caption{\label{f-atomlevels}
Atomic level scheme: two low lying states $|a\rangle$ and $|b\rangle$ are coherently coupled by resonant Raman laser fields with an effective Rabi frequency $\Omega$, and they are separately coupled to two Rydberg states $|r_{1,2}\rangle$ with effective Rabi frequencies $\Omega_{1,2}$. The dots represent a possible population of the atomic levels.
}
\end{figure}

We consider an ensemble of atoms with the level structure shown in Fig.~\ref{f-atomlevels}: two lower states $|a\rangle$ and $|b\rangle$ are coherently coupled to two Rydberg states $|r_{1,2}\rangle$ by laser fields (in most experiments coupling to Rydberg states is achieved by two photon transitions via an intermediate excited atomic state), as well as to each other by a coherent Raman coupling. We assume that the laser fields at different frequencies are derived from a single master or that their phases are otherwise locked to a common reference. Although only the Rydberg coupling between states $|a\rangle$ and $|r_{1}\rangle$ and between $|b\rangle$ and $|r_{2}\rangle$ are shown in Fig.~\ref{f-atomlevels}, we will also make use of coupling between states  $|a\rangle$ and $|r_{2}\rangle$ and between $|b\rangle$ and $|r_{1}\rangle$. There are $N$ atoms in the sample and we assume that their interaction parameters with the fields are identical, so that if initially all the atoms are prepared in the same ground state, say  $|a\rangle$, any action of the fields will lead to states confined to the manifold of symmetric superpositions of all the individual atomic states. For co-propagating laser fields, the propagation phases over the spatial extent of the  ensemble play no role for the ground state coherence, while they may be absorbed in the definition of the the optically excited states to render the field couplings and state amplitudes fully symmetric under permutation of atoms. Let us denote by $|n_a,n_b,n_{r1},n_{r2}\rangle$ the symmetric state in which $n_a$ atoms are in state  $|a\rangle$, $n_b$  are in state  $|b\rangle$, and $n_{r1,r2}$ atoms are in the respective Rydberg state $|r_{1,2}\rangle$.

A single atom of the sample can be laser-excited to a specifically chosen Rydberg state with large principal quantum number $n$\cite{Gallagher}. Due to the strong dipolar coupling between two Rydberg excited atoms, excitation of another atom to the same Rydberg state would not be resonant with the same laser field, but would require a detuning given by the dipole interaction energy. The van der Waals interaction gives rise to frequency shifts of few kHz for $n=45$ Rydberg excited Rb atoms separated by few micrometer distance, while F\"orster resonances occur for special principal quantum numbers and leads to shifts up to few MHz, e.g., for Rb atoms excited to $40p_{3/2}$-states up to 10 micrometers apart \cite{Saffman2010}. This shift is at the root of the Rydberg blockade mechanism: when an ensemble enclosed within a few micon sized volume is uniformly excited with a laser field resonant with a specific Rydberg state, only one excitation occurs within  the ensemble, and as long as the excitation Rabi frequency is well below the  blockade shift, further excitation is energetically blocked. Our Hilbert space is thus restricted to states $|n_a,n_b,n_{r1},n_{r2}\rangle$ where  $n_{r1,r2}$ are confined to 0 and 1 and where $n_a+n_b+n_{r1}+n_{r2}=N$. As can be checked, this restricted Hilbert space has dimension $4N$.

The dynamics between states $|a\rangle$ and $|b\rangle$ can be described by means of the collective spin operator $\vec{J}=\sum_{k}\vec{S_k}$, where the index $k$ denotes individual atoms and the individual atomic spin components are
\begin{eqnarray}
S_x &=& \frac{1}{2}(|a\rangle\langle b| + |b\rangle\langle a|), \\
S_y &=&\frac{i}{2}(-|a\rangle\langle b| + |b\rangle\langle a|), \\
S_z &=&\frac{1}{2}(|a\rangle\langle a| - |b\rangle\langle b|).
\end{eqnarray}
The collective spin components
can be represented by means of creation and annihilation operators as
\begin{eqnarray}
J_x &=& \frac{1}{2}(a^{\dag}b+ab^{\dag}), \\
J_y &=& \frac{i}{2}(a^{\dag}b-ab^{\dag}), \\
J_z &=& \frac{1}{2}(a^{\dag}a-b^{\dag}b),
\end{eqnarray}
where the operator $a$ destroys an atom in state $|a\rangle$, $a^{\dag}$ creates an atom in state $|a\rangle$, (and similarly for $b$), and the commutation relations are $[a,a^{\dag}]=[b,b^{\dag}]=1$.

Hamiltonian terms proportional to the  components of $\vec{J}$ can be realized by electromagnetic fields near resonance with the $a-b$ transition: the detuning of the frequency of the field then gives the $J_z$ contribution and its amplitude $\Omega$ and phase $\varphi$ determine the $J_{x,y}$ contributions to the Hamiltonian.

The single excitation allowed in either of the Rydberg states is distributed in a symmetric fashion as a superposition state with the same excitation amplitude on each individual atom, and the coupling of the  ground state with $n$ atoms to all the components in the excited superposition state leads to an enhancement by the factor $\sqrt{n}$ compared to the single  atom coupling. As the Rydberg excitation reduces the ground state atom number by one, this process couples states of the form $|n,0\rangle$ and $|n-1,1\rangle$, and the $\sqrt{n}$ dependence of the coupling strength is in exact agreement with the matrix element of the Jaynes Cummings Hamiltonian $H_{JC}=ga^+ \sigma_- + g^* a \sigma_-$. The Rydberg excitation of atoms subject to the excitation blockade thus implements the JC Hamiltonian, where the oscillator degree of freedom represents the ground state population and the two-level system represents the Rydberg state population. In our  model, the Hamiltonians that couple the two lower states and the Rydberg states can be described as
\begin{eqnarray}
H_{JC1} &= &\Omega_1 a\sigma^{(1)}_{+}+\Omega_1^{*} a^{\dag}\sigma^{(1)}_{-} ,
\label{HJC1}
\\
H_{JC2} &= &\Omega_2 b\sigma^{(2)}_{+}+\Omega_2^{*} b^{\dag}\sigma^{(2)}_{-},
\end{eqnarray}
where the operators $\sigma^{(1)}_{\pm}$ create or destroy an atom in the Rydberg state $|r_1\rangle$, and similarly for $\sigma^{(2)}_{\pm}$. If the Rydberg coupling field $\Omega_j$ is detuned from resonance by $\delta_{j}$, the Hamiltonian will also contain a term proportional to $\delta_{j}\sigma^{(j)}_{z}$ where  $\sigma^{(j)}_{z}=\sigma^{(j)}_{+} \sigma^{(j)}_{-}-\sigma^{(j)}_{-} \sigma^{(j)}_{+}$.

\section{Dynamic squeezing}
\label{sec-dyn}

As discussed in detail in \cite{Banacloche93}, the JC interaction with a two-level system can evolve coherent states of a harmonic oscillator into  amplitude squeezed states. This is most easily understood if we assume that the two-level system occupies one of the dressed state superpositions,  $(|0\rangle\pm |1\rangle)/\sqrt{2}$, which diagonalize the interaction with a classical resonant field with a real Rabi coupling. The time evolution of a product state of such a dressed state and an initial coherent state of the oscillator is approximately given by a simple phase evolution of each oscillator eigenstate component, corresponding to the Rabi interaction energy, which is proportional to $\sqrt{n}$. This implies a rotation and a twisting of the coherent state amplitude distribution in phase space towards a squeezed state. Unlike the evolution under the usual squeezing Hamiltonian which is quadratic in the creation and annihilation operators, the squeezing in the JC system is not ideal, and the circular phase space distribution does not evolve into an ellipse, but rather into a banana-like shape \cite{Banacloche93}.

To produce a spin-squeezed states in an analogous manner, we initialize the atomic sample with all atoms in an even superposition of states $|a\rangle$ and $|b\rangle$, corresponding to a spin coherent state with maximum eigenvalue of the  $J_x$ operator.
Thus, while each atom is in the superposition $2^{-1/2}(|a\rangle + |b\rangle)$, the collective state is
\begin{eqnarray}
|\psi_0\rangle = 2^{-N/2}\sum_{n_a=0}^{N}\sqrt{N\choose n_a}|n_a, N-n_a,0,0\rangle .
\end{eqnarray}
Next we want, for all values of $n_a$ and $n_b$, within the corresponding binomial distribution, to prepare the two-level dressed states with states $|r_1\rangle$ and $|r_2\rangle$ populated and unpopulated with equal amplitude. Within the quite narrow distribution of $n_{a(b)}$, $\pi/2$ excitation pulses on the two Rydberg transitions, may accomplish this state with adequate precision. The precision may be further enhanced if the Rydberg state is excited by an adiabatic chirp of the Rydberg exciting laser detuning towards resonance. During the chirp, phase factors depending on $n_{a(b)}$ are accumulated, but since their values are known and since they contribute  in the same manner as the phases we shall need for the  squeezing process, we shall assume that they present no problem for the correct initialization of the system.

After having prepared the initial state, the system evolves with the resonant Jaynes-Cummings Hamiltonian $H_{JC}=H_{JC1}+H_{JC2}$, where we choose $\Omega_1=\Omega_2\equiv \Omega_{JC}$. The system thus corresponds to a pair of standard JC systems in which the initial state of each system approximately corresponds to the ``field'' in a coherent state with the mean number of photons $(N-1)/2$ and the ``atom'' in the superposition of the ground and excited states.  Our system differs from the standard JC by two aspects: {\em (i)} the ``photon'' numbers $n_{a,b}$ and the ``atom excitation'' numbers $n_{r1,r2}$ are anticorrelated by $n_a+n_b+n_{r1}+n_{r2}=N$, and  {\em (ii)} having $n_{r1,r2}$ fixed, $n_{a,b}$ have a binomial rather than a Poissonian distribution. The anticorrelation is very important for the shape of the resulting state: whereas the $a$-mode component with $n_a=\bar n+\Delta n$ atoms acquires a phase proportional to $\sqrt{\bar n+\Delta n}\approx \sqrt{\bar n}(1+\frac{\Delta n}{2\bar n}-\frac{\Delta n^2}{8\bar n^2})$, the $b$-mode component with $n_b=\bar n-\Delta n$ atoms acquires a phase proportional to $\sqrt{\bar n-\Delta n}\approx \sqrt{\bar n}(1-\frac{\Delta n}{2\bar n}-\frac{\Delta n^2}{8\bar n^2})$ (assuming the mean excitation number $\bar n = (N-1)/2$). These two contributions add so that the terms linear in $\Delta n$, as well as all other odd power terms, cancel. Apart from an unimportant constant, one is left with a phase evolution proportional to the  even power terms of $\Delta n$, in particular, $\phi \approx -\Omega_{JC}t\Delta n^2/(4\bar n^{3/2})+O(\Delta n^4)$. If terms beyond third order can be neglected (for $\Delta n \ll \bar n$), the evolution in the $a$ and $b$ modes corresponds to that caused by a Hamiltonian $\propto J_z^2$ that leads to spin squeezing by ``one-axis twisting'', as shown in \cite{Kitagawa}. Thus, whereas in the standard JC model the coherent state rotates in the phase space (due to the linear term in $\Delta n$) and aquires a ``banana'' shape during the squeezing procedure, in our system states starting at the equator of the Poincar\'{e} sphere do not rotate and they stay symmetric. Moreover, in comparison with the coherent state with the standard deviation of photon number $=\sqrt{\bar n}$, the binomial distribution is narrower with the standard deviation $=\sqrt{\bar n/2}$. This also contributes to render the higher powers of  $\Delta n$ in the evolution less important than in the standard JC model, and we reach better squeezing than in the standard JC model with the same initial mean excitation number.

\begin{figure}
\centerline{\epsfig{file=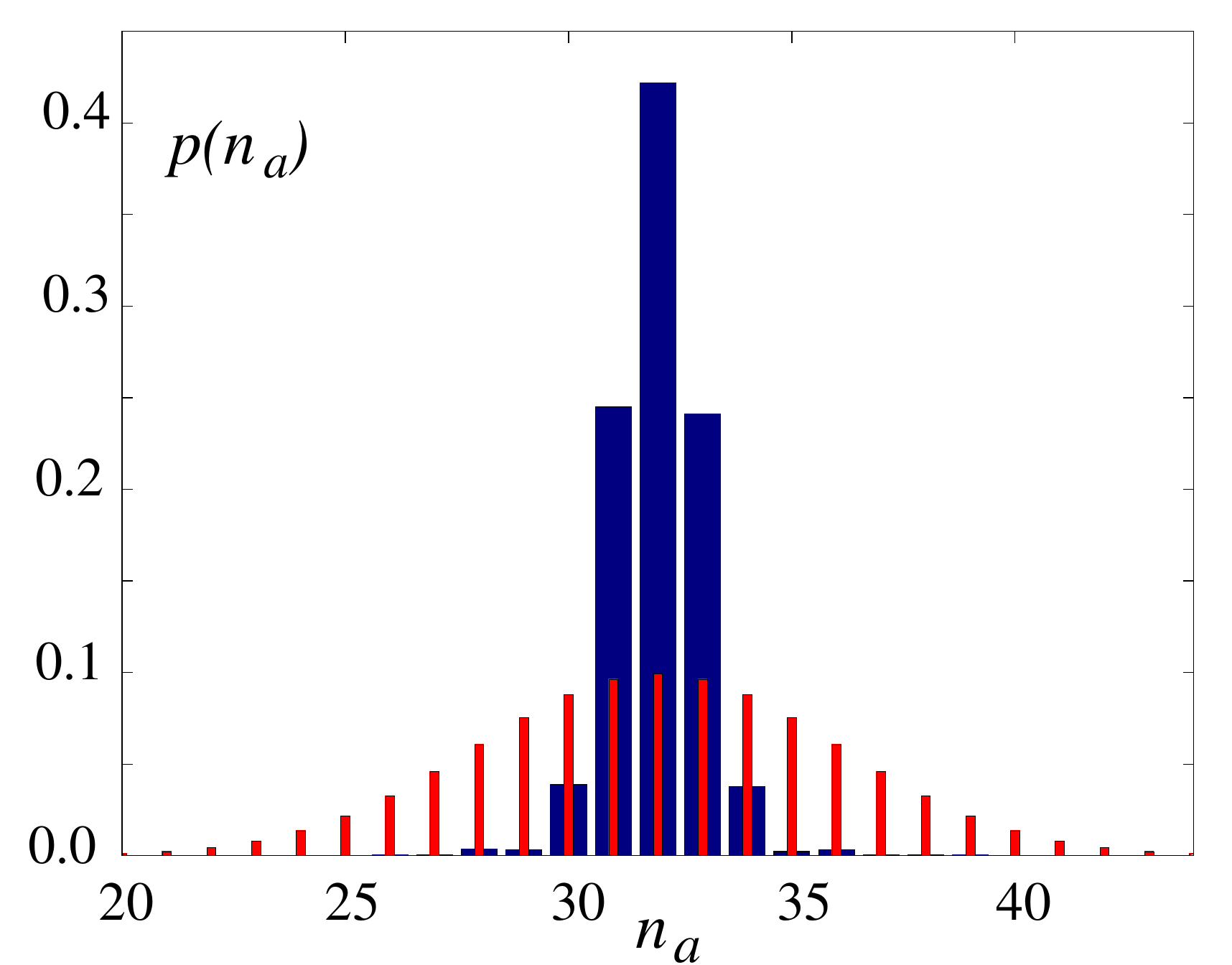,scale=0.4}}
\caption{\label{f-squezstat1}(Color online)
Probability of finding $n_a$ atoms (wide blue bars) in state $|a\rangle$ in a sample of $N=64$ atoms that underwent the dynamical squeezing procedure. The narrow red bars represent statistics of the spin coherent state, for comparison.
}
\end{figure}

The time for reaching the maximum squeezing can be estimated by considering that states with the excitation number around $\bar n+\Delta n$ rotate in the phase space by a phase angle $\sim |\Omega_{\rm JC}|t\Delta n/(2\bar n^{3/2})$ and the initially spin coherent state becomes maximally squeezed when the contributions from the ends of the distribution  $\bar n\pm \sqrt{\bar n/2}$ have rotated to the opposite angles $\pm \pi/2$. Therefore the time of maximum squeezing is $\sim N/|\Omega_{\rm JC}|$. This estimate is well confirmed by our numerical simulations of the process.

The combined JC operators induce a nonzero correlation $\langle J_y J_z + J_z J_y\rangle$ equivalent to a tilted uncertainty ellipse which can be rotated by means of a Raman transfer Hamiltonian $\propto J_x$, so that the two ground state populations acquire subbinomial distributions. The state of the ensemble contains contributions of the Rydberg states, which can be transferred to the ground states by frequency chirping the Rydberg coupling fields out of resonance.  The state is now squeezed in the $J_z$ variable, i.e., it has suppressed fluctuations in the difference of atom numbers $n_a-n_b$ in levels $a$ and $b$.

\begin{figure}
\centerline{\epsfig{file=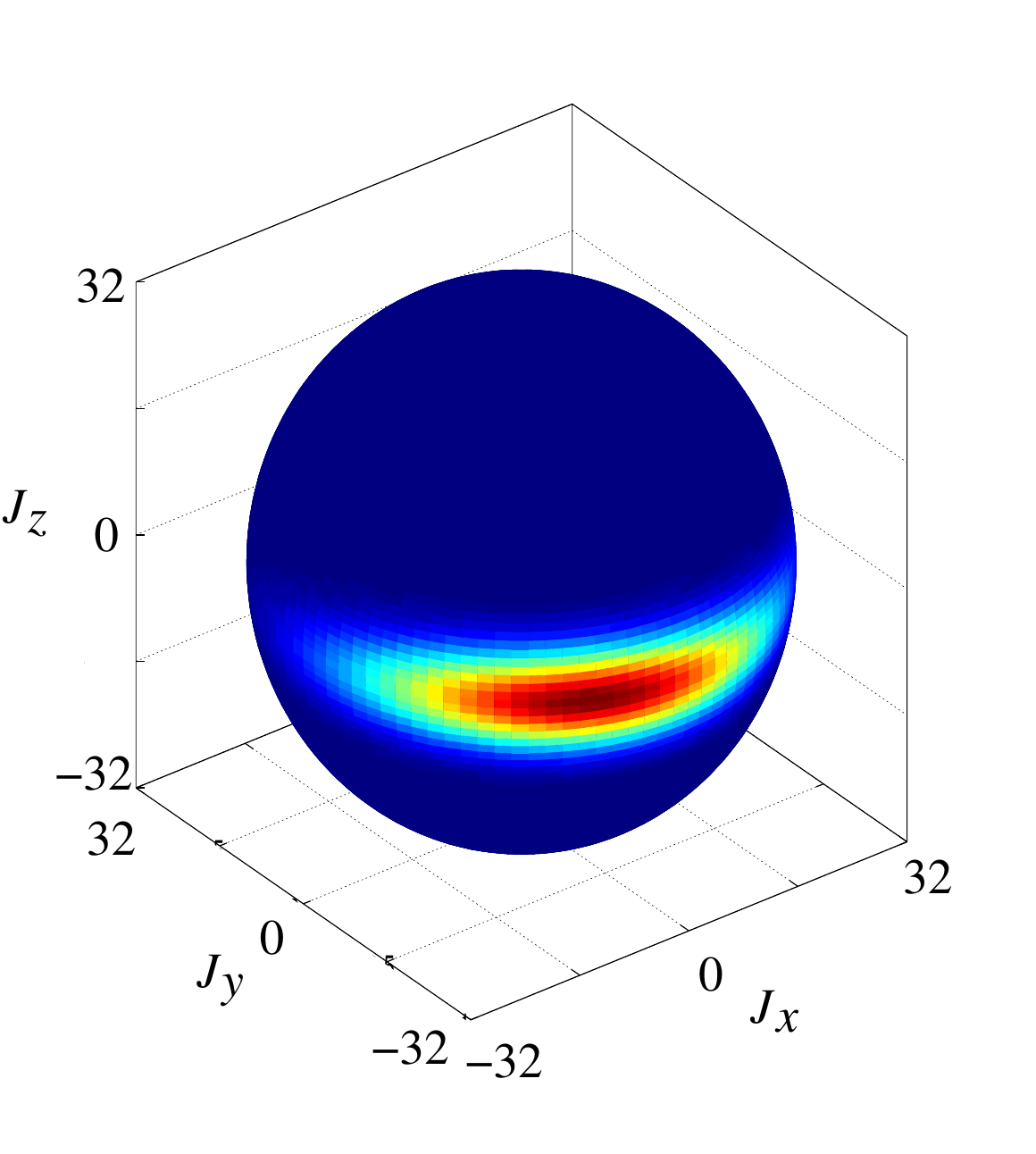,scale=0.5}}
\caption{\label{f-figSqzQp}(Color online)
$Q$-function of the resulting squeezed state in a sample of $N=64$ atoms after the dynamical squeezing procedure.
}
\end{figure}

In Fig.~\ref{f-squezstat1} we show results of our numerical simulation of the procedure for squeezing a state with $N=64$ atoms. The final fluctuation is $\Delta J_z=1.08$ which corresponds to 11 dB squeezing of noise below the fluctuations of a spin coherent state having $\Delta J_z=4$. In Fig.~\ref{f-figSqzQp} the $Q$-function of the resulting state is displayed (the $Q$-function is the trace of the product of the density matrix of the state with the density matrix of the spin coherent state).

\section{Adiabatic squeezing}
\label{sec-adiab}

\subsection{Hamiltonian spectrum}

For an even atom number, $N$, it is in principle possible to reach a collective $J_z=0$ eigenstate with $\Delta J_z=0$,  i.e., $n_a=n_b$, whereas for odd $N$ the maximally squeezed state has $\Delta J_z=1/2$ associated with the two equally populated components $n_a=n_b\pm 1$.
To assess such maximally squeezed states we consider adiabatic transition from an extremal eigenstate of operator $J_x$ (i.e., a spin coherent state) to an extremal eigenstate of $H_{JC}$. Since both of these operators and any weighted combination of them can be implemented with suitable laser fields, we are in principle able to drive this adiabatic transition, and let us show that the extremal eigenstates of $H_{JC}$ are indeed maximally squeezed.

\begin{figure}
\centerline{\epsfig{file=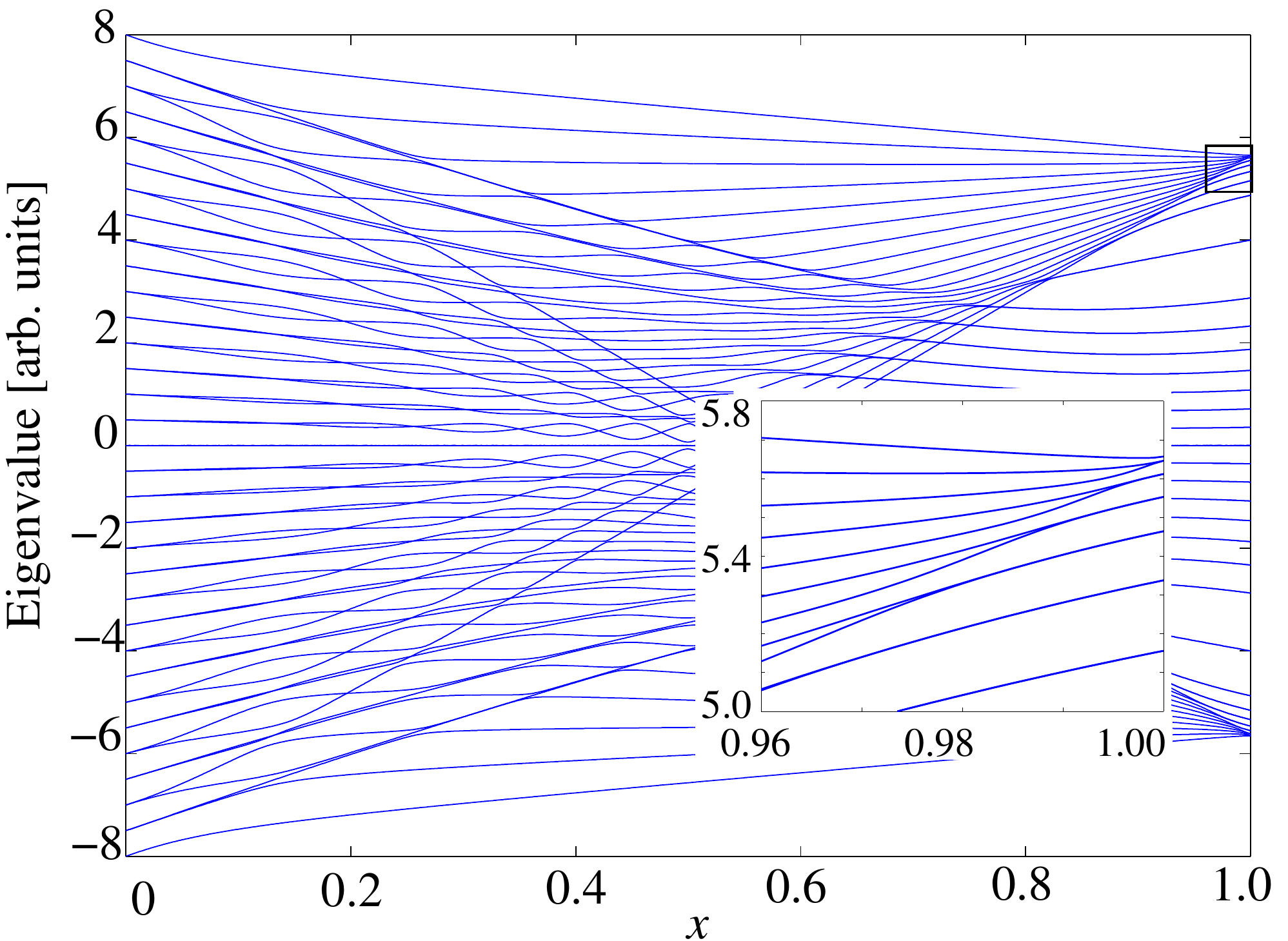,scale=0.45}}
\caption{\label{f-hamdiag}(Color online)
Eigenvalues of Hamiltonian $xH_{JC}+(1-x)J_x$ for $N=16$.
}
\end{figure}

The eigenstates of $H_{JC}$ are dressed states of a form analogous to the standard JC model. Assuming a real coupling strength $\Omega_{JC}$, the usual pair of eigenstates generalizes to four states,
\begin{eqnarray}
|\psi^{(n_a,n_b)}_{+,+} \rangle &=& \frac{1}{2}\left(|n_a,n_b,0,0\rangle + |n_a-1,n_b,1,0\rangle\right.
 \\
& +& \left. |n_a,n_b-1,0,1\rangle + |n_a-1,n_b-1,1,1\rangle\right) ,\nonumber \\
|\psi^{(n_a,n_b)}_{+,-} \rangle &=& \frac{1}{2}\left(|n_a,n_b,0,0\rangle + |n_a-1,n_b,1,0\rangle\right.
 \\
& -& \left. |n_a,n_b-1,0,1\rangle - |n_a-1,n_b-1,1,1\rangle\right) ,\nonumber \\
|\psi^{(n_a,n_b)}_{-,+} \rangle &=& \frac{1}{2}\left(|n_a,n_b,0,0\rangle - |n_a-1,n_b,1,0\rangle\right.
 \\
& +& \left. |n_a,n_b-1,0,1\rangle - |n_a-1,n_b-1,1,1\rangle\right) ,\nonumber \\
|\psi^{(n_a,n_b)}_{-,-} \rangle &=& \frac{1}{2}\left(|n_a,n_b,0,0\rangle - |n_a-1,n_b,1,0\rangle\right.
 \\
& -& \left. |n_a,n_b-1,0,1\rangle + |n_a-1,n_b-1,1,1\rangle\right) ,\nonumber
\end{eqnarray}
with the energies
\begin{eqnarray}
E^{(n_a,n_b)}_{+,+}&=& \Omega_{JC}\left(\sqrt{n_a} + \sqrt{n_b}\right), \\
E^{(n_a,n_b)}_{+,-}&=& \Omega_{JC}\left(\sqrt{n_a} - \sqrt{n_b}\right), \\
E^{(n_a,n_b)}_{-,+}&=& \Omega_{JC}\left(-\sqrt{n_a} + \sqrt{n_b}\right), \\
E^{(n_a,n_b)}_{-,-}&=& \Omega_{JC}\left(-\sqrt{n_a} - \sqrt{n_b}\right).
\end{eqnarray}
For even $N$ the extremal eigenstates have energies $\pm \Omega_{JC}\sqrt{2N}$ and correspond to $n_a=n_b=N/2$, while for odd $N$ each of the extremal states is doubly degenerate, e.g., the states with maximum energy $\Omega_{JC}(\sqrt{(N+1)/2}+\sqrt{(N-1)/2})$ are $|\psi^{((N+1)/2,(N-1)/2)}_{+,+} \rangle$ and $|\psi^{((N-1)/2,(N+1)/2)}_{+,+} \rangle$.
Thus, in the extremal energy eigenstates, the populations of levels $a$ and $b$ are as close to each other as possible.
Eigenvalues of the combined Hamiltonian $xH_{JC}+(1-x)J_x$ for $x$ between 0 and 1 and $N=16$ are shown in Fig.~\ref{f-hamdiag}. As can be seen, an extreme eigenvalue of  $J_x$ is smoothly transformed into the corresponding extreme eigenvalue of  $H_{JC}$ when the parameter $x$ changes between zero and unity. Thus, by properly choosing functions $f_{1,2}(t)$ one can steer the state by adiabatically changing the Hamiltonian  $f_{1}(t)J_x+f_{2}(t)H_{JC}$ such that an initial spin coherent state (i.e., extremal eigenstate of $J_x$) evolves into an  $H_{JC}$ extremal eigenstate. After the adiabatic transfer, one can get rid of the Rydberg excitations by sweeping the Rydberg coupling fields out of resonance,  and if $N$ is even the final state is $|N/2,N/2,0,0\rangle$.
 
If $N$ is odd, the final superposition state, $(|(N+1)/2,(N-1)/2,0,0\rangle+|(N-1)/2,(N+1)/2,0,0\rangle)/\sqrt{2}$, carries a phase that depends on the orientation of the initial phase coherent state. With imballanced JC strengths, one may in principle prepare one of the states, \textit{e.g.}, $J_z=1/2$.

\subsection{Compensation for nonadiabatic transitions}
The adiabatic scenario holds the promise to generate much better squeezing than the dynamic time evolution with the fixed JC interaction, but in the vicinity of the extremal states, $H_{JC}$ has a very narrow level spacing ($\propto \Omega_{JC}N^{-3/2}$). Hence,  too fast parameter changes will couple different adiabatic eigenstates and thus decrease the resulting squeezing. One can deal with this problem by a careful choice of the functions $f_{1,2}(t)$ such that the trade-off between transitions to unwanted states and the speed of the process is optimized. To maintain adiabaticity even for moderate atom numbers will, however, be incompatible with the finite Rydberg state lifetime. We propose here a strategy to actively compensate for the nonadiabatic transitions by additional time-varying Hamiltonian terms. For a given time varying Hamiltonian $H_0(t)$, the well known non-adiabatic coupling terms due to the time dependence of the adiabatic eigenstates are of a form fully equivalent to the application of an extra Hamiltonian in a fixed basis, and as proposed in \cite{Berry}, they can therefore be canceled by applying precisely the negative of that Hamiltonian, which is explicitly given as
\begin{eqnarray}
\label{HamBerry}
H_1(t)=i\sum_{m\neq n}\frac{|m\rangle\langle m|\partial_{t}H_0 |n\rangle \langle n|}{E_n-E_m}.
\end{eqnarray}
A system subject to the combined Hamiltonian  $H_0(t)+H_1(t)$ evolves exactly along the instantaneous eigenstates of  $H_0(t)$.
It is of course in general not easy to provide exactly the Hamiltonian $H_1(t)$ from physically available interactions. Even though one may not be able to construct the full Hamiltonian $H_1(t)$, one can, however, attempt to preserve as close as possible the evolution of the extremal adiabatic eigenstate $|\psi_0(t)\rangle$ of $H_0(t)$ by the application of a judiciously chosen pertubation of the system.  Among the operators that can be naturally implemented, and which couple the adiabatic states so that they can be used to counter the non-adiabatic transitions, are
\begin{eqnarray}
 H_{JC}^{(y)}&= & i\left( a\sigma^{(1)}_{+}- a^{\dag}\sigma^{(1)}_{-}
+ b\sigma^{(2)}_{+}- b^{\dag}\sigma^{(2)}_{-} \right), \\
 H_{JC}^{(y,{\rm cross})}&= & i\left( b\sigma^{(1)}_{+}- b^{\dag}\sigma^{(1)}_{-}  + a\sigma^{(2)}_{+}- a^{\dag}\sigma^{(2)}_{-} \right).
\end{eqnarray}
Note that $H_{JC}^{(y)}$ has a similar form as $H_{JC}$ (uses the same transitions between the lower levels and the Rydberg states), but the phases of the coupling fields are different. The operator $H_{JC}^{(y,{\rm cross})}$ uses fields driving ``cross''-transitions between states $a$ and $r_2$ and between states $b$ and $r_2$ (not shown in Fig.~\ref{f-atomlevels}). To partially compensate for the nonadiabatic transitions, we thus suggest to drive the system with the Hamiltonian
\begin{eqnarray}
 \label{eqHamCompens}
 H(t)=H_0(t)+\alpha_1(t) H_{JC}^{(y)}+\alpha_2(t)H_{JC}^{(y,{\rm cross})},
\end{eqnarray}
where
\begin{eqnarray}
 \label{eqHamAdiabat}
 H_0(t)=f_{1}(t)J_x+f_{2}(t)H_{JC}
\end{eqnarray}
is the original Hamiltonian and the functions $\alpha_{1,2}$ are chosen such that the norm of the vector $(\alpha_1(t) H_{JC}^{(y)}+\alpha_2(t)H_{JC}^{(y,{\rm cross})}-H_1(t))|\psi_0(t)\rangle$ is minimized. This results in an  explicit formula for $\alpha_{1,2}$ involving only matrix elements of the operators $H_{JC}^{(y)}$, $ H_{JC}^{(y,{\rm cross})}$ and $H_1(t)$ in the state $|\psi_0(t)\rangle$ (see Appendix \ref{ap-alphas}).

We note that the availability of a number of different interaction Hamiltonians with variable strengths is the prerequisite of optimal control theory \cite{Control1,Control2}, and that strong numerical methods exist to identify the fastest and most reliable route towards desired final states. Our approach towards a useful choice of parameters is almost with certainty not the optimal one. We believe, however, that it offers an interesting, fast and explicit protocol, and that it retains a physical interpretation, which guides our efforts to choose the few, most relevant interaction terms.

\subsection{Results}
\begin{figure}
\centerline{\epsfig{file=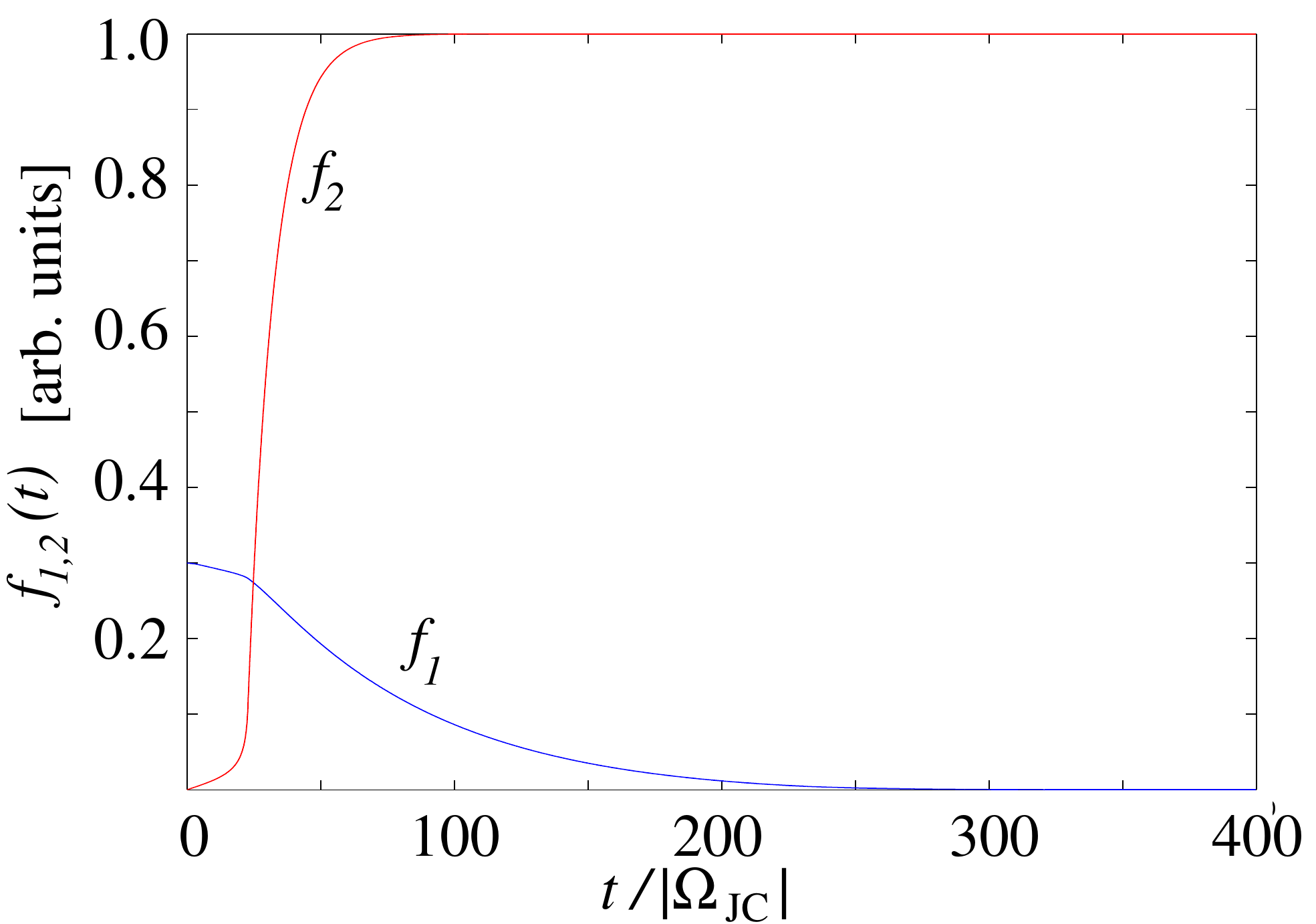,scale=0.45}}
\caption{\label{f-HamParams}(Color online)
Evolution of parameters $f_{1}(t)$ and $f_{2}(t)$ in the Hamiltonian $H_0(t)$ chosen such as to minimize nonadiabatic transitions provided the maximum JC coupling is limited and the state should proceed to the final stage as fast as possible. The time is given in units $1/|\Omega_{JC}|$ corresponding to the maximum JC coupling. The number of atoms is $N=15$.
}
\end{figure}

\begin{figure}
\centerline{\epsfig{file=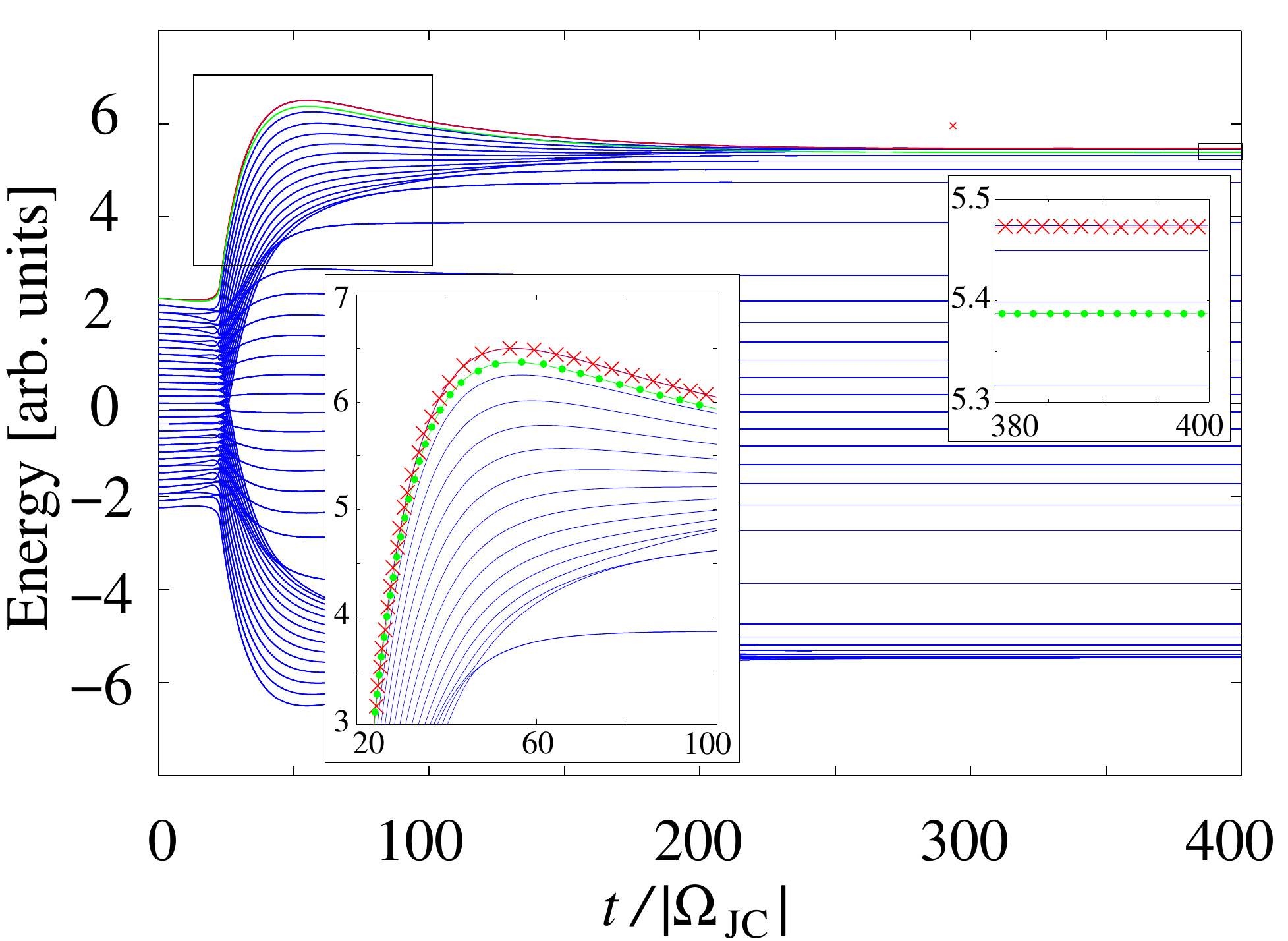,scale=0.45}}
\caption{\label{f-hamdiag-time}(Color online)
Time evolution of the spectrum of Hamiltonian  $H_0(t)$ with parameters as in Fig.\ref{f-HamParams}. The red curve (close to top, marked with ``x'' in detail) is the mean value of energy calculated during the evolution.  For comparison, the green curve (marked with dots in detail) is the mean value of energy for the procedure without the compensation of nonadiabatic transitions.
}
\end{figure}

\begin{figure}
\centerline{\epsfig{file=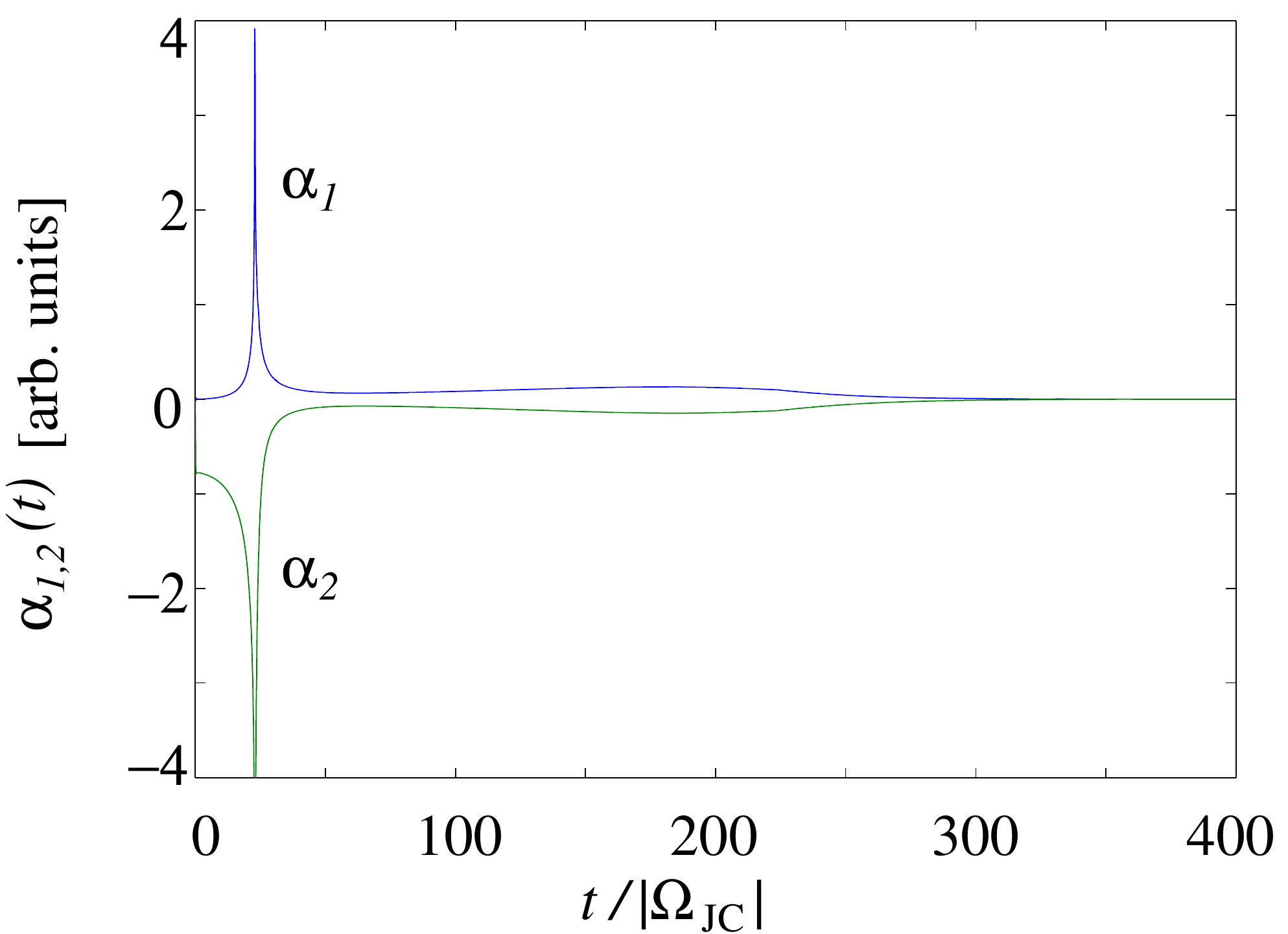,scale=0.45}}
\caption{\label{f-BerryParam}(Color online)
Time dependence of parameters $\alpha_{1,2}$ of the Hamiltonian Eq.~(\ref{eqHamCompens}) chosen such as to minimize the nonadiabatic transitions.
}
\end{figure}

\begin{figure}
\centerline{\epsfig{file=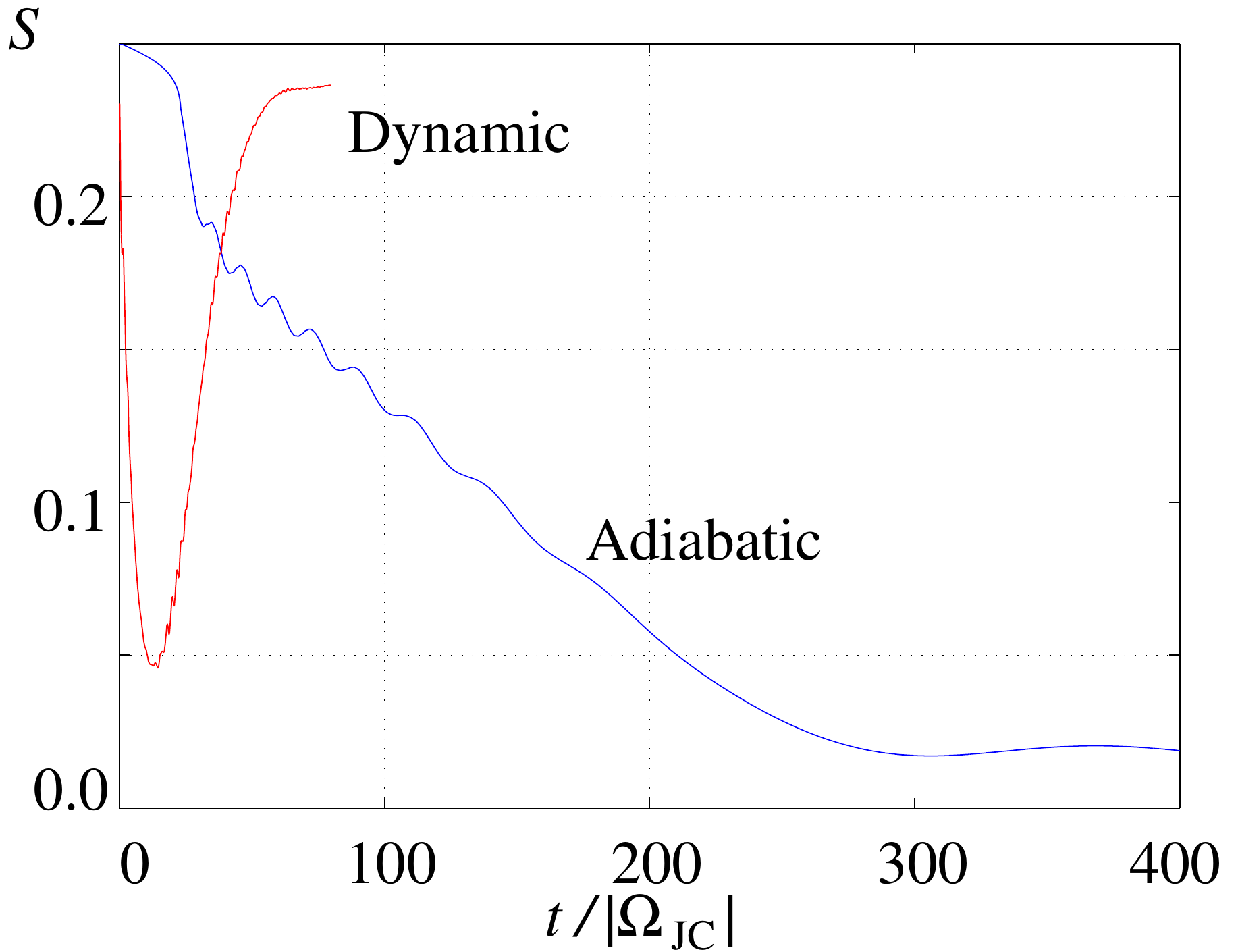,scale=0.45}}
\caption{\label{f-timevolvSQZ}(Color online)
Time evolution of the squeezing parameter $S$ in the dynamic regime and in the adiabatic procedure with compensation of non-adiabatic transitions, cf., Eq.~(\ref{eqHamCompens}).
}
\end{figure}

The procedure and its results are shown in Figs.~\ref{f-HamParams}--\ref{f-timevolvSQZ} for a system with $N=15$ atoms. The rate of the transition increases with the magnitude of parameters $f_{1,2}$. The fastest process would occur if one switches the JC coupling $f_2(t) H_{JC}$ to the maximum possible value and then slowly turns off the $J_x$ part of the Hamiltonian, the rate of turning  $J_x$  off increases with the magnitude of $f_2$. Since the JC coupling to the Rydberg state cannot be arbitrarily strong (because one may, e.g., excite more than one atom to the Rydberg state if $\Omega_{JC}$ becomes comparable to the blockade splitting), it is natural to choose the maximum allowed value of the JC coupling as the principal restriction in the optimization procedure. In our very simple scenario, the algorithm in each time step chooses the change of parameters  $f_{1,2}$ such as to approach the target values ($f_1\to 0, f_2\to 1$, see Fig.~\ref{f-HamParams}) as fast as possible, while having the amplitudes of nonadiabatic transitions within a predetermined tolerance interval.

The resulting time dependent spectrum of the Hamiltonian $H_0(t)$ is shown in Fig.~\ref{f-hamdiag-time}. The slow approach to the terminal state is necessary due to the tight level spacing of $H_{JC}$. The figure also shows as a green (red) curve the mean value $\langle H_0(t)\rangle$ in the evolving state without (with)  compensation for the nonadiabatic transitions according to Eq.~(\ref{eqHamCompens}). The perfect agreement of the red curve and the extremal eigenvalue of $H_0(t)$ demonstrates that our compensation significantly improves the adiabatic following. 

In Fig.~\ref{f-BerryParam} we show the time dependence of the compensation parameters $\alpha_{1,2}$ of Eq.~(\ref{eqHamCompens}). It turns out that the available operators only have substantial overlap with the Berry compensation Hamiltonian $H_1$ of Eq.~(\ref{HamBerry}) in the initial stage when $H_0$ is dominated by $J_x$. In the later stages this overlap becomes very small and the compensation becomes virtually inefficient. However, even with such a limited option for compensation the additional Hamiltonians enable us to perform the transition substantially faster, by a factor of $\sim$5--10.

 We may terminate the process described in Figs.~\ref{f-hamdiag},~\ref{f-BerryParam} at any instant and apply a chirp of the JC interactions to remove the Rydberg excitation. The resulting time dependent value of the squeezing parameter $S = \langle \Delta J_z^2 \rangle/N$ is shown in Fig.~\ref{f-timevolvSQZ}. Note that a phase coherent state with $\langle J_z \rangle = 0$ (i.e., our initial state) has $S = 1/4$ corresponding to the binomial distribution, and any state having $S < 1/4$ is squeezed. As can be seen, the dynamical procedure squeezes the state faster but after reaching a certain minimum value the parameter $S$ returns to values of very poor squeezing. The adiabatic process works more slowly but leads to a much deeper squeezing.

\section{Schr\"{o}dinger cat generation}
\label{sec-schroed}

To generate superpositions of coherent spin states with opposite spin directions, one starts with the same initial state as in the preceding sections, but uses coupling only to one of the Rydberg levels. The scheme has some similarities to that of standard JC generation of Schr\"{o}dinger cats discussed in \cite{Buzek1992}, but new interesting features stem from the spherical topology of the phase space and of the finite Hilbert space of the atomic sample. The coherent state superpositions are interesting not only per se as highly non-classical states, but were discussed as, e.g., a resource for building quantum logic gates \cite{Ralph03,Fiurasek10}.

Let us assume that the Hamiltonian is $H_{JC1}$ of Eq. (\ref{HJC1}) with $\Omega_1=\Omega_1^{*}=\Omega_{JC}$ and let the Hamiltonian be switched on for time $\tau$, where
\begin{eqnarray}
\tau = \left\{ \begin{array}{lr}
\frac{\pi}{\Omega_{JC}}\sqrt{\frac{N}{2}} & {\rm for} \ N \ {\rm even}
\label{tau1} \\
\frac{\pi}{\Omega_{JC}}\sqrt{\frac{N-1}{2}} &  {\rm for} \ N \ {\rm odd}
\end{array}
\right. .
\end{eqnarray}
During this time the basis states of $J_z$ and their superpositions evolve as described in Appendix \ref{ap-superp}.
For smoothly changing coefficients $a_{\Delta n} \approx a_{\Delta n+1}$ of the superposition the total superposition contains pairs
of states with one atom being either in the Rydberg state or in state $|b\rangle$. One can eliminate the Rydberg excitation by a pulse coupling the atomic states $|b\rangle$ and $|r_1\rangle$ of duration $\tau_0$ with
\begin{eqnarray}
\tau_0 = \left\{ \begin{array}{lr}
\frac{\pi}{\Omega_{JC}}\sqrt{\frac{2}{N}} & {\rm for} \ N \ {\rm even} \\
\frac{\pi}{\Omega_{JC}}\sqrt{\frac{2}{N-1}} &  {\rm for} \ N \ {\rm odd}
\end{array}
\right. .
\end{eqnarray}
After that, the state contains almost exclusively contributions with no Rydberg excitation and with even $\Delta n$. The contribution of odd $\Delta n$ states is suppressed to the order $\Delta n^2/N$ which is small for coherent states with  $\Delta n$ limited to $\Delta n \lesssim \sqrt{N}/2$. 
In particular, starting with a spin coherent state with $N=20$, the resulting state has 92\% contribution of even $\Delta n$ and 8\%  of odd $\Delta n$.
Thus, one has created a state that is analogous to the even superposition of coherent states with oposite amplitudes.

\begin{figure}
\centerline{\epsfig{file=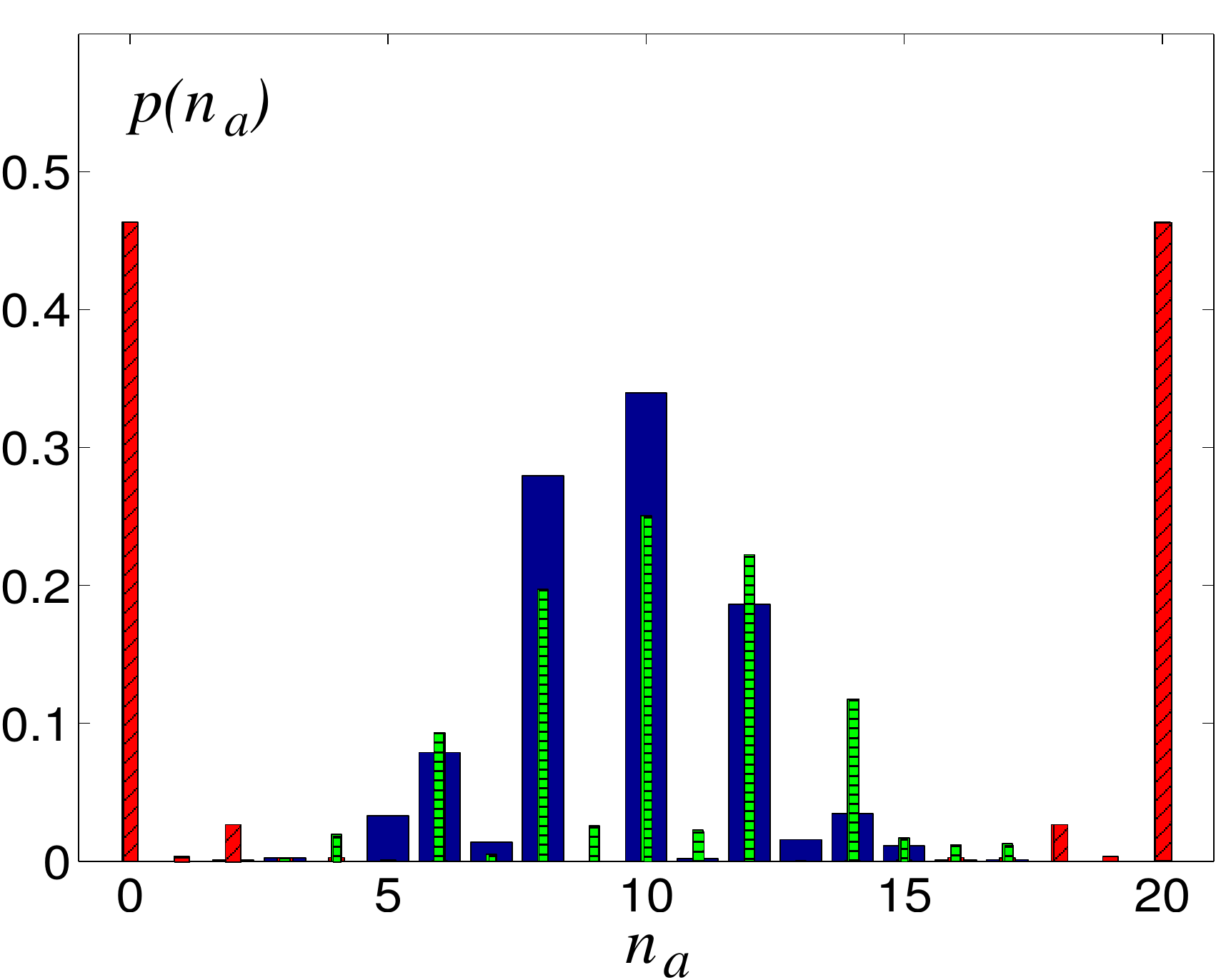,scale=0.4}}
\caption{\label{f-figCat1}(Color online)
Population of state $|a\rangle$ (wide blue bars) after the procedure of Schr\"{o}dinger cat generation, with $N=20$. The narrow red bars with tilted hatching show the population after rotation of the state by $\exp(i J_x \pi/2)$ showing the two separate peaks (live and dead cat). The narrow green bars with horizontal hatching show the population after rotation of the state by $\exp(i J_y \pi/2)$.}
\end{figure}

\begin{figure}
\centerline{\epsfig{file=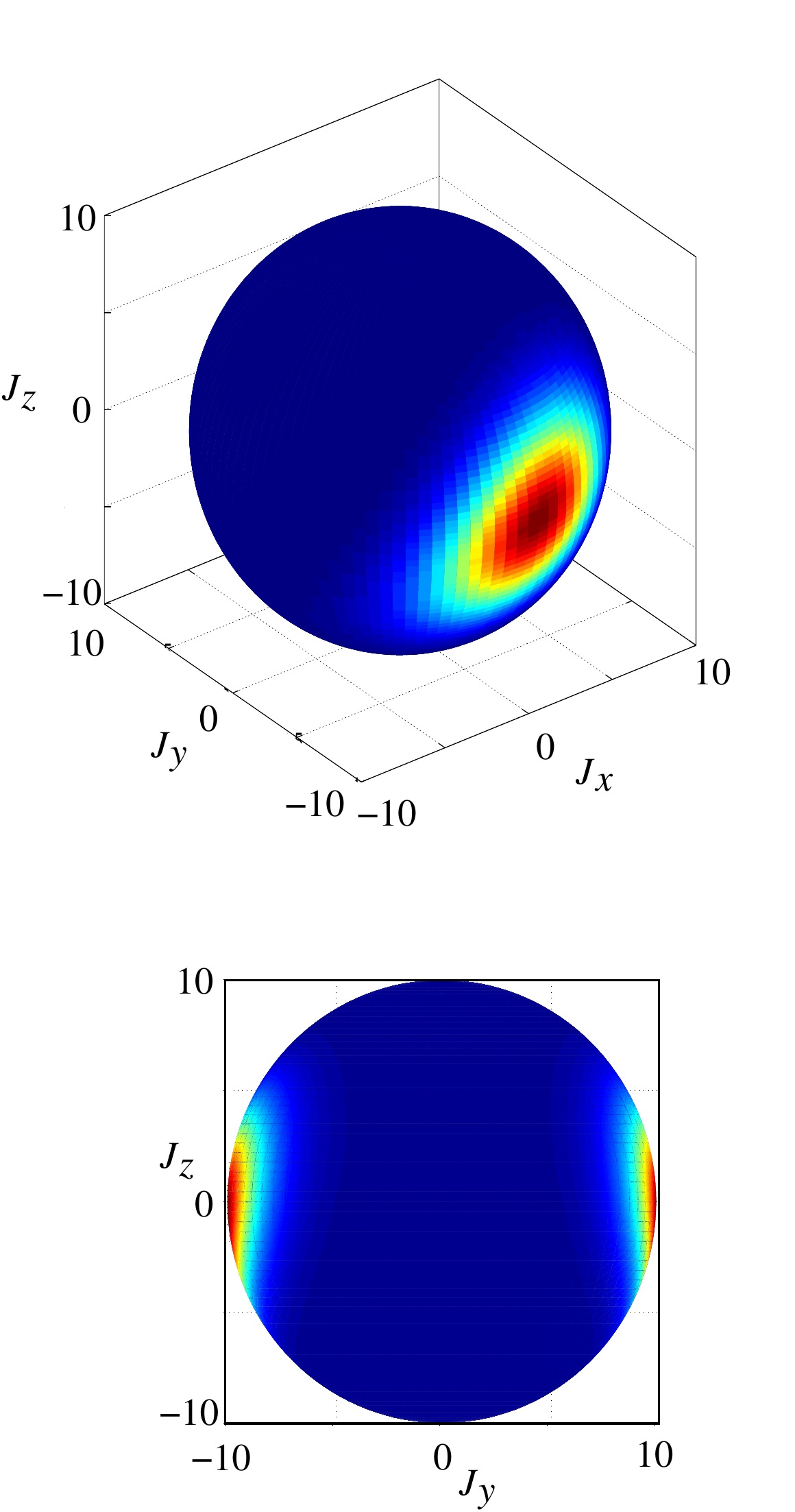,scale=0.45}}
\caption{\label{f-figCat3}(Color online)
$Q$-function of the resulting state after the Schr\"{o}dinger cat generation procedure: two peaks on opposite sides of the sphere are produced.}
\end{figure}

\begin{figure}
\centerline{\epsfig{file=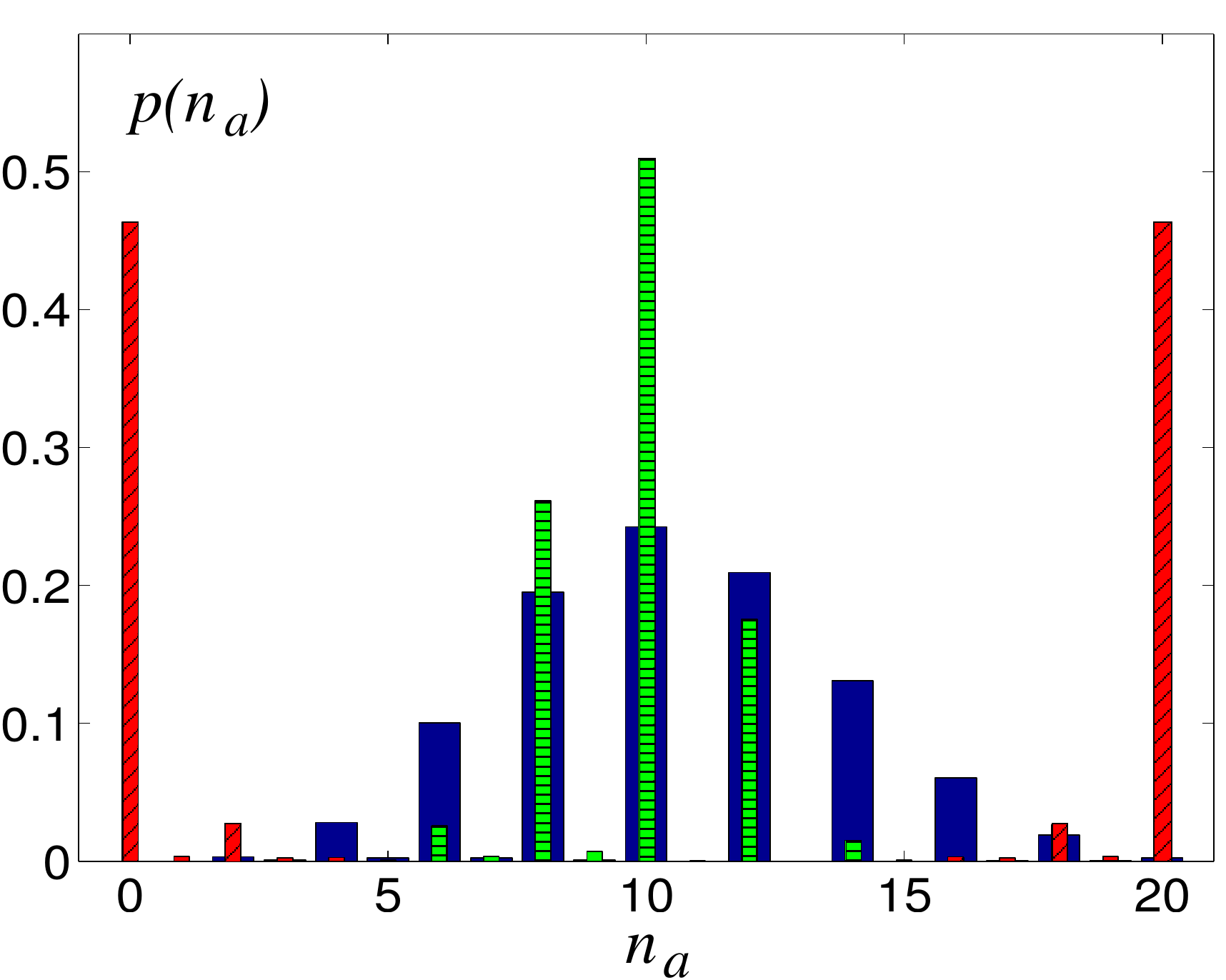,scale=0.4}}
\caption{\label{f-figCat2}(Color online)
Same as in Fig. \ref{f-figCat1} after the rotation of the state by $\exp(i J_y \phi)$ where the angle $\phi$ is optimized to have the statistics squeezed (narrow green bars with vertical hatching) or anti-squeezed (wide blue bars).}
\end{figure}

One can observe analogies of the most important features of the Schr\"{o}dinger cat states familiar from optics: two separate peaks in one quadrature, interference fringes in the complementary quadrature, and ``photon number'' oscilations in the Fock representation.
The results of our numerical simulation for $N=20$ can be seen in Figs. \ref{f-figCat1}--\ref{f-figCat2}. Although during the procedure the Rydberg state is populated with $\sim 50\%$ probability, after the last step the remaining population of the Rydberg state is only $1.1 \%$. In Fig. \ref{f-figCat1} one can see the oscillation in the atomic population statistics (blue bars) analogous to the photon number oscillation in the even superposition of coherent states with opposite amplitudes. The red bars are the atomic population statistics after rotation of the resulting state by $\exp(i J_x \pi/2)$. This is analogous to detecting an optical Schr\"{o}dinger cat  state by means of the homodyne scheme with the phase chosen to maximally separate the peaks. The green bars correspond to the state after rotation by $\exp(i J_y \pi/2)$. In optics this would correspond to observing the state by homodyning such that the two peaks completely overlap and one detects the interference fringes.

A $Q$-function of the resulting state can be seen in Fig. \ref{f-figCat3}. Only one component of the superposition is visible in the upper part of the figure,  
as the other one is on the opposite side of the sphere. One can see that the state is squeezed. To observe the squeezing by measuring the population of the atomic states, one can rotate the resulting state by  $\exp(i J_y \phi)$ with a properly chosen angle $\phi$. The resulting statistics is shown in Fig.  \ref{f-figCat2}. In this case one can see the green-bar statistics being narrower and the blue-bar statistics wider. In optical analogy this corresponds to a superposition of two phase-squeezed coherent states with super Poissonian statistics.  In this case the contribution of odd $\Delta n$ of the resulting state dropped to 1.1\%.

\section{Conclusion}
\label{sec-concl}
We have proposed a scheme that takes advantage of the Rydberg blockade in a spin polarized atomic sample to employ features of the JC model in a new environment. The finite atom number and the possibility to switch on and off the JC coupling between various atomic states makes the spectrum of interesting effects rather broad. In comparison to the standard JC model, one can squeeze coherent states symetrically, avoiding the ``banana-like'' deformation. By adiabatically changing the Hamiltonian from the angular-momentum operator to the double JC Hamiltonian, one can achieve full squeezing. With a proper choice of the JC Hamiltonian, one can prepare Schr\"{o}dinger cat-like superpositions of spin polarized states with opposite spins. These procedures can be useful, e.g., for precision time measurements, or for quantum information processing.

\acknowledgments
T.O. was supported by the Czech Science Foundation, grant No. GAP205/10/1657. K. M. was supported by the EU Integrated Project AQUTE.

\appendix
\begin{widetext}
\section{Parameters for the nonadiabatic transitions compensation}
\label{ap-alphas}
Taking the norm of the vector $(\alpha_1 H_{JC}^{(y)}+\alpha_2 H_{JC}^{(y,{\rm cross})}-H_1)|\psi_0(t)\rangle$ and expressing it as a function $F(\alpha_1, \alpha_2)$ of (real) parameters $\alpha_{1,2}$ one gets
\begin{eqnarray}
F(\alpha_1, \alpha_2) &=& \langle \psi_0|(\alpha_1 H_{JC}^{(y)}+\alpha_2 H_{JC}^{(y,{\rm cross})}-H_1)^2
|\psi_0 \rangle \nonumber \\
 &=& \langle \psi_0|H_1^2 |\psi_0\rangle + \alpha_1^2 \langle \psi_0| H_{JC}^{(y)2} |\psi_0\rangle
+ \alpha_2^2 \langle \psi_0| H_{JC}^{(y,{\rm cross})2} |\psi_0\rangle
- \alpha_1 \langle \psi_0| (H_{JC}^{(y)} H_1 +H_1 H_{JC}^{(y)})|\psi_0\rangle \nonumber \\
& & - \alpha_2 \langle \psi_0| (H_{JC}^{(y,{\rm cross})} H_1 +H_1 H_{JC}^{(y,{\rm cross})})|\psi_0\rangle
+ \alpha_1 \alpha_2 \langle \psi_0| (H_{JC}^{(y,{\rm cross})}H_{JC}^{(y)}  +H_{JC}^{(y)} H_{JC}^{(y,{\rm cross})})|\psi_0\rangle .
\end{eqnarray}
Requiring for the extremum $\partial F(\alpha_1, \alpha_2)/\partial \alpha_1 = \partial F(\alpha_1, \alpha_2)/\partial \alpha_2 = 0$ one gets two equations for the unknowns $\alpha_{1,2}$ with the solution
\begin{eqnarray}
\alpha_1 &=& \frac{2\langle H_{JC}^{(y,{\rm cross})2}\rangle\langle H_{JC}^{(y)} H_1 +H_1 H_{JC}^{(y)}\rangle - \langle H_{JC}^{(y,{\rm cross})}H_{JC}^{(y)}  +H_{JC}^{(y)} H_{JC}^{(y,{\rm cross})}\rangle \langle H_{JC}^{(y,{\rm cross})}H_1  +H_1 H_{JC}^{(y,{\rm cross})}\rangle
}{4\langle H_{JC}^{(y)2}\rangle\langle H_{JC}^{(y,{\rm cross})2}\rangle - \langle H_{JC}^{(y,{\rm cross})}H_{JC}^{(y)}  +H_{JC}^{(y)} H_{JC}^{(y,{\rm cross})}\rangle ^2 } ,
\\
\alpha_2 &=& \frac{2\langle H_{JC}^{(y)2}\rangle\langle H_{JC}^{(y,{\rm cross})} H_1 +H_1 H_{JC}^{(y,{\rm cross})}\rangle - \langle H_{JC}^{(y,{\rm cross})}H_{JC}^{(y)}  +H_{JC}^{(y)} H_{JC}^{(y,{\rm cross})}\rangle \langle H_{JC}^{(y)}H_1  +H_1 H_{JC}^{(y)}\rangle
}{4\langle H_{JC}^{(y)2}\rangle\langle H_{JC}^{(y,{\rm cross})2}\rangle - \langle H_{JC}^{(y,{\rm cross})}H_{JC}^{(y)}  +H_{JC}^{(y)} H_{JC}^{(y,{\rm cross})}\rangle ^2 } ,
\end{eqnarray}
where the mean value $\langle \dots \rangle$ is taken in state $|\psi_0(t)\rangle$.


\section{Superposition generation for Schr\"{o}dinger cat states}
\label{ap-superp}
Let us assume Hamiltonian  $H_{JC1}$ of Eq. (\ref{HJC1}) with $\Omega_1=\Omega_1^{*}=\Omega_{JC}$  being switched on for time $\tau$ of Eq. (\ref{tau1}).
For even $N$, an initial state of the form
$ \left| \frac{N}{2}+\Delta n, \frac{N}{2}-\Delta n, 0 \right\rangle $
evolves as (up to the second order in $\Delta n$)
\begin{eqnarray}
\left| \frac{N}{2}+\Delta n, \frac{N}{2}-\Delta n, 0 \right\rangle \to
(-1)^{(N+\Delta n)/2}\left( \left| \frac{N}{2}+\Delta n, \frac{N}{2}-\Delta n, 0 \right\rangle
+ i \frac{\Delta n^2 \pi}{4 N}\left| \frac{N}{2}+\Delta n-1, \frac{N}{2}-\Delta n, 1 \right\rangle
\right)
\end{eqnarray}
for  $\Delta n$ even, and
\begin{eqnarray}
\left| \frac{N}{2}+\Delta n, \frac{N}{2}-\Delta n, 0 \right\rangle \to
(-1)^{(N+\Delta n-1)/2}\left(  \frac{\Delta n^2 \pi}{4 N}  \left| \frac{N}{2}+\Delta n, \frac{N}{2}-\Delta n, 0 \right\rangle
- i \left| \frac{N}{2}+\Delta n-1, \frac{N}{2}-\Delta n, 1 \right\rangle
\right)
\end{eqnarray}
for  $\Delta n$ odd.
For odd $N$, an initial state of the form
$ \left| \frac{N-1}{2}+\Delta n, \frac{N-1}{2}-\Delta n, 0 \right\rangle $
evolves as
\begin{eqnarray}
\left| \frac{N-1}{2}+\Delta n, \frac{N-1}{2}-\Delta n, 0 \right\rangle \to
\nonumber \\
(-1)^{\frac{N-1+\Delta n}{2}}\left( \left| \frac{N-1}{2}+\Delta n, \frac{N-1}{2}-\Delta n, 0 \right\rangle
+ i \frac{\Delta n^2 \pi}{4 (N-1)}\left| \frac{N-1}{2}+\Delta n-1, \frac{N-1}{2}-\Delta n, 1 \right\rangle
\right)
\end{eqnarray}
for  $\Delta n$ even, and
\begin{eqnarray}
\left| \frac{N-1}{2}+\Delta n, \frac{N-1}{2}-\Delta n, 0 \right\rangle \to
\nonumber \\
(-1)^{\frac{N+\Delta n}{2}}\left(  -\frac{\Delta n^2 \pi}{4 (N-1)}  \left| \frac{N-1}{2}+\Delta n, \frac{N-1}{2}-\Delta n, 0 \right\rangle
+ i \left| \frac{N-1}{2}+\Delta n-1, \frac{N-1}{2}-\Delta n, 1 \right\rangle
\right)
\end{eqnarray}
for  $\Delta n$ odd.
Thus, for even $N$ a superposition of the form
\begin{eqnarray}
|\Psi_0\rangle = \sum_{\Delta n} a_{\Delta n}\left| \frac{N}{2}+\Delta n, \frac{N}{2}-\Delta n, 0 \right\rangle
\end{eqnarray}
evolves into
\begin{eqnarray}
|\Psi_{\tau}\rangle = (-1)^{N/2}
\left\{ \sum_{\Delta n \ {\rm even}}
(-1)^{\Delta n/2}
\times \left[ a_{\Delta n}\left| \frac{N}{2}+\Delta n, \frac{N}{2}-\Delta n, 0 \right\rangle
- i  a_{\Delta n+1}\left| \frac{N}{2}+\Delta n, \frac{N}{2}-\Delta n -1, 1 \right\rangle
\right. \right.
\nonumber \\
\left. \left.
+ \frac{(\Delta n+1)^2\pi}{4N}a_{\Delta n+1}
\left| \frac{N}{2}+\Delta n+1, \frac{N}{2}-\Delta n-1, 0 \right\rangle
-i \frac{(\Delta n+2)^2\pi}{4N}a_{\Delta n+2}
\left| \frac{N}{2}+\Delta n+1, \frac{N}{2}-\Delta n-2, 1 \right\rangle
\right] \right\} ,
\end{eqnarray}
and for odd $N$
a superposition of the form
\begin{eqnarray}
|\Psi_0\rangle = \sum_{\Delta n} a_{\Delta n}\left| \frac{N-1}{2}+\Delta n, \frac{N-1}{2}-\Delta n, 0 \right\rangle
\end{eqnarray}
evolves into
\begin{eqnarray}
|\Psi_{\tau}\rangle = (-1)^{\frac{N-1}{2}}
\left\{ \sum_{\Delta n \ {\rm even}}
(-1)^{\Delta n/2}
\times \left[ a_{\Delta n}\left| \frac{N-1}{2}+\Delta n, \frac{N-1}{2}-\Delta n, 0 \right\rangle
 \right.\right.
\nonumber \\
\left. \left.
- i  a_{\Delta n+1}\left| \frac{N-1}{2}+\Delta n, \frac{N-1}{2}-\Delta n -1, 1 \right\rangle
+ \frac{(\Delta n+1)^2\pi}{4(N-1)}a_{\Delta n+1}
\times
\left| \frac{N-1}{2}+\Delta n+1, \frac{N-1}{2}-\Delta n-1, 0 \right\rangle
\right. \right.
\nonumber \\
\left. \left.
-i \frac{(\Delta n+2)^2\pi}{4(N-1)}a_{\Delta n+2}  
\times
\left| \frac{N-1}{2}+\Delta n+1, \frac{N-1}{2}-\Delta n-2, 1 \right\rangle
\right] \right\} .
\end{eqnarray}
\end{widetext}

\end{document}